\documentclass[letterpaper]{article}
\usepackage{chicagor}
\usepackage{aaai1}
\usepackage{times}
\usepackage{helvet}
\usepackage{courier}
\usepackage{times}
\usepackage{chicagor}

\newcommand{\complex}{\C}
\newcommand{\out}{{{\cal O}}}
\newcommand{\complexs}{{\bf C}}
\newcommand{\bF}{{\bf F}}

\newtheorem{THEOREM}{Theorem}[section]
\newenvironment{theorem}{\begin{THEOREM}}%
                        {\end{THEOREM}}
\newtheorem{LEMMA}[THEOREM]{Lemma}
\newenvironment{lemma}{\begin{LEMMA}}%
                      {\end{LEMMA}}
\newtheorem{COROLLARY}[THEOREM]{Corollary}
\newenvironment{corollary}{\begin{COROLLARY}}%
                          {\end{COROLLARY}}
\newtheorem{PROPOSITION}[THEOREM]{Proposition}
\newenvironment{proposition}{\begin{PROPOSITION}}%
                            {\end{PROPOSITION}}
\newtheorem{DEFINITION}[THEOREM]{Definition}
\newenvironment{definition}{\begin{DEFINITION}}%
                            {\end{DEFINITION}}
\newtheorem{CLAIM}[THEOREM]{Claim}
\newenvironment{claim}{\begin{CLAIM}}%
                            {\end{CLAIM}}
\newtheorem{EXAMPLE}[THEOREM]{Example}
\newenvironment{example}{\begin{EXAMPLE}}%
                            {\end{EXAMPLE}}
\newtheorem{REMARK}[THEOREM]{Remark}
\newenvironment{remark}{\begin{REMARK}}%
                            {\end{REMARK}}

\newcommand{\thm}{\begin{theorem}}
\newcommand{\lem}{\begin{lemma}}
\newcommand{\pro}{\begin{proposition}}
\newcommand{\dfn}{\begin{definition}}
\newcommand{\rem}{\begin{remark}}
\newcommand{\xam}{\begin{example}}
\newcommand{\cor}{\begin{corollary}}
\newcommand{\prf}{\noindent{\bf Proof:} }
\newcommand{\ethm}{\end{theorem}}
\newcommand{\elem}{\end{lemma}}
\newcommand{\epro}{\end{proposition}}
\newcommand{\edfn}{\bbox\end{definition}}
\newcommand{\erem}{\bbox\end{remark}}
\newcommand{\exam}{\bbox\end{example}}
\newcommand{\ecor}{\end{corollary}}
\newcommand{\eprf}{\bbox\vspace{0.1in}}
\newcommand{\beqn}{\begin{equation}}
\newcommand{\eeqn}{\end{equation}}

\newcommand{\bbox}{\vrule height7pt width4pt depth1pt}

\newcommand{\clm}{\begin{claim}}
\newcommand{\eclm}{\end{claim}}

\newcommand{\rarrow}{\rightarrow}

\newcommand{\IR}{\mbox{$I\!\!R$}}

\newcommand{\IN}{\mbox{$I\!\!N$}}

\renewcommand{\phi}{\varphi}

\newcommand{\C}{{\cal C}}
\newcommand{\D}{{\cal D}}

\newcommand{\F}{{\cal F}}

\newcommand{\M}{{\cal M}}

\newcommand{\W}{{\cal W}}

\newcommand{\<}{\langle}
\renewcommand{\>}{\rangle}

\newcommand{\ol}{\setlength{\itemsep}{0pt}\begin{enumerate}}
\newcommand{\eol}{\end{enumerate}\setlength{\itemsep}{-\parsep}}
\newcommand{\ul}{\setlength{\itemsep}{0pt}\begin{itemize}}
\newcommand{\dl}{\setlength{\itemsep}{0pt}\begin{description}}
\newcommand{\edl}{\end{description}\setlength{\itemsep}{-\parsep}}
\newcommand{\eul}{\end{itemize}\setlength{\itemsep}{-\parsep}}

\newcommand{\commentout}[1]{}

\newcommand{\bi}{\begin{itemize}}
\newcommand{\ei}{\end{itemize}}
\newcommand{\be}{\begin{enumerate}}
\newcommand{\ee}{\end{enumerate}}

\newcommand{\ZK}{{\mathcal{ZK}}}
\newcommand{\bit}{\ensuremath{\{0,1\}}}
\renewcommand{\W}{I}

\def\full{0} 
\ifnum\full=1
	\newcommand{\fullv}[1]{#1}
	\newcommand{\shortv}{\commentout}
\else
	\newcommand{\shortv}[1]{#1}
	\newcommand{\fullv}{\commentout}
\fi

\begin{document}

\title{I Don't Want to Think About it Now:\\Decision
  Theory With Costly Computation}
\author{
Joseph Y. Halpern%
\thanks{Supported in part by NSF grants IIS-0534064, IIS-0812045, and
IIS-0911036, and by AFOSR grants 
FA9550-08-1-0438 and FA9550-09-1-0266, and ARO grant W911NF-09-1-0281.} 
\and Rafael Pass\thanks{
Supported in part by a Microsoft New Faculty Fellowship, 
NSF CAREER Award CCF-0746990, AFOSR Award
FA9550-08-1-0197, and BSF Grant 2006317.}\\ 
Cornell University\\
\mbox{$\{$halpern $\mid$ rafael$\}$@cs.cornell.edu}} 
\date{ }
\maketitle

\begin{abstract}
Computation plays a major role in decision
making.  Even if an agent is willing to ascribe a probability to all states
and a utility to all outcomes, and maximize expected utility, doing so
might present serious computational problems.  Moreover, computing the
outcome of a given act might be difficult.  In a companion paper we
develop a framework for game theory  with costly computation, where the
objects of choice are Turing machines.  Here we apply that framework to
decision theory.  We show how well-known phenomena like 
\emph{first-impression-matters biases} (i.e., 
people tend to put more weight on evidence they hear early on), 
\emph{belief polarization} (two people
with different prior beliefs, hearing the same evidence, can end up with
diametrically opposed conclusions), and 
the \emph{status quo} bias
(people are much more likely to stick with what they already have)   
can be easily captured in that framework.  Finally, we 
use the framework  to define some new notions: \emph{value of computational
information} (a computational variant of \emph{value of information})
and \emph{computational value of conversation}.
\end{abstract}

\section{Introduction}
Computation plays a major role in decision
making.  Even if an agent is willing to ascribe a probability to all states
and a utility to all outcomes, and maximize expected utility---that is,
to follow the standard prescription of rationality as 
recommended by Savage \citeyear{Savage}, doing so might present serious
computational problems.  Computing the relevant probabilities might be
difficult, as might computing the relevant utilities.  Work on
Bayesian networks \cite{Pearl} and other representations of probability,
and related work on representing utilities \cite{BG95,BBDHP04}
can be viewed as attempts to ameliorate these computational problems.
Our focus is on the complexity of computing the outcome of an act in a
given state. Consider the following simple example, taken from 
\cite{HP10}.

Suppose that a decision maker (DM) is given an input $n$, and is asked
whether it is prime.  The DM gets a payoff of \$1,000 if he gives the
correct answer and loses \$1,000 if he gives the wrong answer.  However,
he also has the option of playing safe, and saying ``pass'', in which
case he gets a payoff of \$1.  Clearly, many DMs would say ``pass'' on
all but simple inputs, where the answer is obvious,
although what counts as a ``simple'' input may depend on the DM.%
\footnote{While primality testing is now known to be in polynomial time
\cite{AKS02}, and there are computationally-efficient randomized
algorithms that that give the correct answer with extremely high
probability \cite{Rabin80,SolovayS77}, we can assume that the DM has no
access to a computer.}  

In \cite{HP10}, we introduced a model of game theory
with costly computation.  Here we apply that framework to decision
theory.  We assume that the DM can be viewed as choosing an algorithm
(i.e., a Turing machine); with each Turing machine (TM) $M$ and input, we
associate its \emph{complexity}.   The complexity can represent, for
example, the running time of $M$ on that input, the space used,
the complexity of $M$ (e.g., how many states it has), or the difficulty of
finding $M$ (some algorithms are more obvious than others).  We
deliberately keep the complexity function abstract, to allow for the
possibility of representing a number of different intuitions.  The DM's
utility can then depend, not just on the payoff, but on the complexity.

The DM's goal is to choose the ``best'' TM; the one that will give him
the greatest expected utility, taking both the payoff and complexity
into account.  To make this choice, the DM must have beliefs about the
TM's running time and the ``goodness'' of the TM's output.  For example,
if the TM outputs ``prime'' on some input $n$, then TM must have beliefs
about how likely $n$ is to actually be prime.  As this example suggests,
we actually need here to deal with what philosophers have called
``impossible'' possible worlds \cite{Hi2,Rant}.  If $n$ is a prime, then
this is a mathematical fact; there can be no state where $n$ is not
prime; nevertheless, since we want to allow for DMs that are
resource-bounded and cannot compute whether $n$ is prime, we want it to
be possible for the DM to believe that $n$ is not prime.  Similarly, if
the complexity function is supposed to measure running time, then  the
actual running time of a TM $M$ on input $t$ is a fact of mathematics;
nevertheless, we want to allow the DM to have false beliefs about $M$'s
running time.  
We capture such false beliefs by having both the utility function and the
complexity function depend on the state of nature.

As we show here, using these simple ideas leads to quite a powerful 
framework.  For example, many concerns expressed by the emerging
field of \emph{behavioral economics} (pioneered by Kahneman and 
Tversky \cite{KST82}) can be accounted for by simple assumptions
about players' 
cost of computation.  To illustrate this point, we show that
\emph{first-impression-matters biases} \cite{mrabin98}, that is, that
people tend to put more weight on evidence they hear early on, can be
easily captured using computational assumptions.  We can similarly
explain \emph{belief polarization} \cite{LRL79}---that two people,
hearing the same 
information (but with possibly different prior beliefs) can end up with
diametrically opposed conclusions.
Finally, we can also use the framework to formalize one of the
intuitions for the well-known \emph{status quo} bias
\cite{SZ88}: people are much more likely to stick with what they
already have.  

\commentout{
For example, Samuelson and Zeckhauser \citeyear{SZ88}
observed that when Harvard University professors were offered the
possibility of enrolling in some new health-care options, older
faculty, who were already enrolled in a plan, enrolled in the new option
much less often than new faculty.  Assuming that all faculty evaluate
the plans in essentially the same way, this can be viewed as an instance
of a status-quo bias.  Samuelson and Zeckhauser suggested a number of
explanations for this phenomenon, one of which was computational.  As
they point out, the choice to undertake a careful analysis of the
options is itself a decision.  Someone who is already enrolled in a plan
and is relatively happy with it can rationally decide that it is not
worth the cost of analysis (and thus just stick with her current plan),
while someone who is not yet enrolled is more likely to decide that the
analysis is worthwhile.  We have nothing new to add to this explanation,
beyond the observation that it can be readily modeled in our framework.   
}

As a final application, we use the framework  to define a new notion:
\emph{value of computational information}.  To explain it, we first recall 
\emph{value of information}, a standard notion in decision analysis.
Value of information is meant to be a measure of how much a DM
should be  willing to 
pay to receive new information.  The idea is that, before receiving the
information, the DM has a probability on a set of relevant events and
chooses the action that maximizes his expected utility, given that
probability.  If he receives new information, he can update his
probabilities (by conditioning on the information) and again choose the
action that maximizes expected utility.  The difference between the
expected utility before and after receiving the information is the value
of the information.

In many cases, a DM seems to be receiving valuable information that
is not about what seem to be the relevant events.  This means that we
cannot do a value of computation calculation, at least not in the
obvious way.  For example, suppose that the DM is interested in
learning a secret, which we assume for simplicity is a number between 1
and 1000.  A priori, suppose that the DM takes each number to be
equally likely, and so has probability $.001$.   
Learning the secret has utility, say, \$1,000,000; not learning it has
utility 0.  The number is locked in
a safe, whose combination is a 40-digit binary numbers.  
What is the value to the DM of learning the first 20 digits of the
combination?  As far as value of information goes, it seems that the
value is 0.  The events relevant to the expected utility are the possible
values of the secret;
learning the combination does not change the probabilities of
the numbers at all.  This is true even if we put the possible
combinations of the lock into the sample space.  
%
On the other hand, it is clear that people may well be willing to pay
for learning the first 20 digits.  It converts an infeasible
problem (trying $2^{40}$ combinations by brute force) to a feasible
problem (trying $2^{20}$ combinations).  

Although this example is clearly contrived, there are many far more
realistic situations where people are clearly 
willing to pay for information to improve computation.  For example,
companies pay to learn about a manufacturing process that will speed up
production; people buy books on speedreading; and faster algorithms for
search clearly are considered valuable.
We show that we can use our computational framework to make the notion
of \emph{value of computational information} precise, in a way that
makes it a special case of value of information.%
\footnote{Our notion of value of computational information is related
to, but not quite the same as, the notion of \emph{value of computation}
introduced by Horvitz \citeyear{Hor87,Hor01}; see Section~\ref{sec:discuss}.}
In addition, we define a notion of
\emph{computational value of conversation}, where the DM can communicate 
interactively with an informed observer before making a decision (as
opposed to just getting some information).
Interestingly, the notion of \emph{zero knowledge} \cite{GMR} gets an
elegant interpretation in this framework.  Roughly speaking, a
zero-knowledge algorithm for 
membership in a language $L$ is one where there is no added value of
conversation in running the algorithm beyond what there would be in
learning whether an input $x$ is in $L$, no matter what random variable
is of interest to the DM.

In the next section we define our computational framework carefully, and
show how it delivers reasonable results in a number of examples.  In
Section~\ref{sec:voci}, we consider the value of computational
information.  We conclude with a discussion of related work in
Section~\ref{sec:discuss}.

\section{A computational framework}\label{sec:framework}
The framework we use here for adding computation to decision theory is
essentially a single-agent version of what were called in \cite{HP10}
\emph{Bayesian machine games}.  In a standard Bayesian game, each player
has a \emph{type} in some set $T$, and then makes a single move.  
Player $i$'s type can be viewed as describing $i$'s
initial information; some facts that $i$ knows about the world.  
In the number-in-the-safe example, there is essentially only one
type, since the DM gets no information.  In the case of the
manufacturing process, the type could be the configuration of the
system; manufacturing processes typically apply to a number of
configurations. We assume that an agent's move consists of choosing a
Turing machine.  As we said in the introduction, associated with each
Turing machine and type is its complexity.   Given as input a type, the
Turing machine outputs an action.  The utility of a player depends on
the type \emph{profile} (i.e., the types of all the players), the
action profile, and the complexity profile.  (While typically all that
matters to player $i$ is the complexity of his algorithm, it may, for
example, matter to him that his algorithm is faster than that of player
$j$.) 

Turning to decision theory, 
we take a \emph{standard decision problem with types} to be characterized by a
tuple $(S,T,A,\Pr,u)$, where $S$ is a state space, $T$ is a set of types,
$A$ is a set of actions, $\Pr$ is a probability distribution on $S
\times T$ (there may be correlation between states and types), 
and $u: S \times T \times A \rightarrow \IR$, where $u(s,t,a)$ is the
DM's utility if he performs action $a$ in state $s$ and has type $t$.%
\footnote{In \cite{HP10}, we did not have a state space $S$, but we
assumed that nature had a type.  Nature's type can be identified with
the state.}  (It is not typical to consider a decision maker's type in
standard decision theory, but it does not hurt to add it; it will prove
useful once we consider computation.)
For each action $a$,
we can consider the 
random variable $u_a$ defined on $S$ by taking $u_a(s,t) = u(s,t,a)$.
The expected utility of action $a$, denoted $E_{\Pr}[u_a]$, 
is just the expected value of the random variable $u_a$ with respect to
the probability distribution $\Pr$; 
that is,  $E_{\Pr}[u_a] = \sum_{(s,t) \in S \times T} \Pr(s,t)u(s,t,a)$.
We assume that 
the DM is an expected utility maximizer, so he chooses an action $a$
with the largest expected utility.

To combine the ideas of Bayesian machine games and decision problems, we
consider \emph{computational decision problems}.  In a computational
decision problem, just like in a computational Bayesian machine game,
the DM chooses a Turing machine.  
We assume that 
the action performed by the TM
depends on the type.
We denote by $M(t)$ the output of the machine on input the type $t$.
To capture the DM's uncertainty about the TM's output, we use an {\em
output function} $\out : {\bf M} \times S \times T   
\rightarrow \IN$, where ${\bf M}$ denotes the set of Turing Machines;
$\out(M,s,t)$ is used to describe what the DM thinks the output of $M(t)$ is in state $s$.
To simplify the presentation, we abuse notation and use $M(s,t)$ to denote
$\out(M,s,t)$. 

The DM's utility will depend on the state $s$, his type $t$, and the
action $M(s,t)$, as is standard; in addition, it will depend on the
``complexity'' of $M$ given input $t$.
The complexity of a machine can represent, for example, the 
running time or space usage of $M$, or the complexity of $M$ itself, or
some combination of these factors.  For
example, Rubinstein \citeyear{Rub85} considers what can be viewed as 
special case of our model, where the DM chooses a finite automaton (and
has no type); the complexity of $M$ is the number of states in the
description of the automaton. 
To capture the cost of computation formally, we use a \emph{complexity
function} $\complex: {\bf M} \times S \times T  
\rightarrow \IN$, 
to describe the complexity of a TM given an input type and state.
(As we shall see, by allowing the state to be included as an argument to
$\complex$, we can capture the DM's uncertainty about the complexity.)

We define a \emph{computational decision problem} to be a tuple
$\D = (S,T,A,\Pr,\M,\complex, \out, u)$, where $S$, $T$, $A$, and $\Pr$
are as in the definition of a standard decision problem,
$\M \subseteq {\bf M}$ is a set of TMs (intuitively, the set that 
the DM can choose among),  
$\out$ is an output function,
$\complex$ is a complexity measure, and $u: S \times T \times A \times \IN
\rarrow \IR$.  The expected utility of a TM $M$
\fullv{in the decision problem $\D$, denoted $U_{\D}(M)$, is}
\shortv{in the decision problem $\D$ is}
$\sum_{s \in S, t \in T} \Pr(s,t) u(s,t,\out(M,s,t),\complex(M,s,t))$.
Note that now the utility function gets the complexity of $M$ as an
argument.  
For ease of exposition here, we restrict to deterministic TMs for most
of the paper; we need to consider randomized TMs for our results on zero
knowledge.  

\commentout{
[[I THINK WE SAY MOVE THE PROPERTY FUNCTION MATERIAL TO SECTION 4,
ALTHOUGH I HAVEN'T DONE THIS.]]
In many cases of interest, the utility of an outcome depends on
whether the action $M(s,t)$ satisfies some property that depends, in
general, on $s$ and $t$ as well.  Formally, define a \emph{property
function} to be a function $f: S
\times T \times \M \rightarrow \IN$; say that a utility function is
dependent on the property function $f$ and a \emph{property utility
function} $u': \IN \times
\IN \rightarrow \IR$ if 
$u(s,t,M(t),\complex(s,t,M)) = u'(f(s,t,M),\complex(s,t,M))$ for all
$s$, $t$, and $M$.  We say that
$u$ is a \emph{property-dependent utility} if it depends on some
$f$ and $u'$.  In the remainder of the paper, we
restrict to property-dependent utility functions.  This is without loss
of generality, since, using standard techniques, we can  
encode $s$, $t$, and $M$ in $f(s,t,M)$ (so that $s$, $t$, and $M$ can be
reconstructed from $f(s,t,M)$).   
The following two examples illustrate property-dependent utilities.
}

\xam\label{pd1}  \emph{Consider the primality-testing problem discussed in 
the introduction.  Formally, suppose that the
DM's type is just a natural number $< 2^{40}$, and the DM must determine
whether the type is prime.  The DM can  choose either 0 (the number
is not prime), 1 (the number is prime), or 2 (pass).  
If $M$ is a TM, then $M(s,t)$ is $M$'s output in state $s$ on input $t$.
The state $s$ here is used to capture the DM's uncertainty about the
output.  So if the DM believes that the DM will output pass with
probability $2/3$, then the set of states such that $M(s,t) = 2$ has
probability $2/3$.
Let
$\complex(s,t,M)$ be 0 if $M$ computes the answer within $2^{20}$
steps on input $t$, and 10 otherwise.  
(Think of $2^{20}$ steps as representing representing a hard deadline.)
Here the state $s$ encodes the DM's 
uncertainty about the running time of $M$.  For example, if the DM does not
know the running time of $M$, but ascribes probability $2/3$ to $M$
finishing in less than $2^{20}$ steps on input $t$, then the set of
states $s$ such that $\C(s,t,M) = 0$ has probability $2/3$.
Finally, let utility $u(s,t,a,c) = 10 - c$ if $a$ is either 0 or 1, and
this is the correct answer in state $s$ (that is, $t$ is viewed as prime
in state $s$ and $a=1$, or $t$ is not viewed as prime in state $s$ and $a=0$),
and $u(s,t,2,c) = 1-c$.  Now the 
state $s$ is used to encode the DM's uncertainty about the correctness
of $M$'s answer.
(Note that we are allowing ``impossible'' states, where $t$ is viewed as
prime in state $s$ even though it is in fact composite; this is needed
to model the DM's uncertainty.)
Thus, if the DM is sure that $M$ always gives
the correct output, then $u(s,t,a,c) = 10 - c$ for all states $s$ and $a
\in \{0,1\}$.
}

\emph{
We can also consider a variant
of this problem, where the DM is given a specific input $t$ and is asked if
$t$ is prime.  Although there is obviously a right answer (the number
is prime or it's not), the DM might still have uncertainty regarding
whether a particular TM $M$ gives the right answer, the running time of
$M$, and the output of $M$.
}
\exam

\commentout{
The reader may wonder why the property function $f$ is a function of
$s$, $t$, and $M$, rather 
than just $s$, $t$, and $a$.  Intuitively, this is because the DM may be
uncertain about the ``goodness'' of an algorithm.  Suppose that 
$M$ and $M'$ are two TMs, and the DM is certain that $M$ correctly
determines primality, but is uncertain about $M'$.  Suppose that in fact
$M(t) = M'(t)$.  If we had used a function $f'$ that depended only on
$s$, $t$, and $a$, then we would have $f'(s,t,M(t)) = f'(s,t,M'(t))$; we
want to allow the DM to consider it possible that $f$ is such that
$f(s,t,M) \ne f(s,t,M')$.   This will become clearer in the next section.
}

\xam\label{pd2}  \emph{Consider the number-in-the-safe example from the
introduction.  Here there is only a single type, $t_0$; we can think of
the state space $S$ as consisting of pairs $(s_1, s_2,s_3)$, where $s_1$ is
the number in the safe, $s_2$ is the combination, and $s_3$ encodes the
DM's beliefs about the complexity and correctness of TMs.
An algorithm in this case is just a sequence of combinations to try and
a stopping rule.   Suppose that the agent gets utility $10 - 
\complex((s_1,s_2,s_3),t_0,M)$ if $s_2$ (the actual combination) is one of 
the numbers 
generated by $M$ before it halts, and $0 - \complex((s_1,s_2,s_3),t_0,M)$
otherwise, 
where $\complex((s_1,s_2,s_3),t_0,M)$ 
is 0 if $M$ halts within $2^{20}$ steps in state $(s_1,s_2,s_3)$, and 10
otherwise. 
}
\exam

\xam
[Biases in information processing]
\label{informationbias:ex} 
{\rm Psychologists have observed many systematic biases in the way that
individuals update their beliefs as new information is received (see
\cite{mrabin98} for a survey). In particular, a
``first-impressions-matter'' bias has been observed: individuals put too
much weight on initial signals and less weight on later signals. As they
become more convinced that their beliefs are correct, many individuals
even seem to simply ignore all information once they reach a confidence
threshold. 
Several papers in behavioral economics have focused on
identifying and modeling some of these biases (see, e.g., \cite{mrabin98}
and the references therein, \cite{M98}, and \cite{RS99}). In
particular, Mullainathan \citeyear{M98} makes a potential connection
between memory and biased 
information processing, using a model that makes several explicit
(psychology-based) assumptions on the memory process (e.g., that the
agent's ability to recall a past event depends on how often he has
recalled the event in the past). 
More recently, Wilson \citeyear{W02} has presented an elegant model of bounded
rationality, where agents are described by finite automata, which
(among other things) can explain why agents eventually choose to ignore
new information; her analysis, however, is very complex and holds only
in the limit 
(specifically, in the limit as the probability $\nu$ that a given round
is the last round goes to 0).

As we now show, the first-impression-matters bias can be easily
explained if we assume that there is a small cost for ``absorbing'' new
information. 
Consider the following simple game (which is very similar to the one
studied by Mullainathan \citeyear{M98} and Wilson \citeyear{W02}). 
The state of nature is a bit $b$ that is $1$ 
with probability $1/2$. An agent receives as his type a sequence of
independent samples $s_1, s_2,\ldots, s_n$ where $s_i = b$ with
probability $\rho > 1/2$. The samples corresponds to signals the agents
receive about $b$.  
An agent is supposed to output a guess $b'$ for the bit $b$. If the
guess is correct, he receives $1-mc$ as utility, and $-mc$ otherwise,
where $m$ is the number of bits of the type he read, and $c$ is the cost
of reading a single bit ($c$ should be thought of the cost of
absorbing/interpreting information). 
It seems
reasonable to assume that $c>0$; signals usually require some effort to
decode (such as reading a newspaper article, or attentively watching a
movie). 
If $c>0$, it easily follows by the Chernoff bound that after
reading a certain (fixed) number of signals $s_1, \ldots, s_i$, the
agents will have a sufficiently good estimate of $\rho$ that the
marginal cost of reading one extra signal $s_{i+1}$ is higher than the
expected gain of finding out the value of $s_{i+1}$. 
That is, after processing a certain number of signals, agents will
eventually  disregard all future signals and base their output 
guess only on the initial sequence. 
We omit the straightforward details.

Essentially the same approach allows us to capture belief polarization.
Suppose for simplicity that two agents start out with slightly different
beliefs regarding the value of some random variable $X$ (think of $X$ as
representing something like ``O.J. Simpson is guilty''), and get the
same sequence $s_1, s_2, \ldots, s_n$  of evidence regarding the value
of $X$.  (Thus, now the type consists of the initial belief, which can
for example be modeled as a probability or a sequence of evidence
received earlier, and the new sequence of evidence).  Both agents update
their beliefs by conditioning.  As before, 
there is a cost of processing a piece of evidence, so once a DM gets
sufficient evidence for either $X=0$ or $X=1$, he will stop processing
any further evidence.  If the initial evidence supports $X=0$, but the
later evidence supports $X=1$ even more strongly, the agent
that was initially inclined towards $X=0$ may raise his beliefs to be
above threshold, and thus stop
processing, believing that $X=0$, while the agent initially 
inclined towards $X=1$ will continue processing and eventually
believe that $X=1$.
}
\exam

\xam[Status quo bias]
{\rm
The status quo bias is well known.  To take just one example,
Samuelson and Zeckhauser \citeyear{SZ88}
observed that when Harvard University professors were offered the
possibility of enrolling in some new health-care options, older
faculty, who were already enrolled in a plan, enrolled in the new option
much less often than new faculty.  Assuming that all faculty evaluate
the plans in essentially the same way, this can be viewed as an instance
of a status quo bias.  Samuelson and Zeckhauser suggested a number of
explanations for this phenomenon, one of which was computational.  As
they point out, the choice to undertake a careful analysis of the
options is itself a decision.  Someone who is already enrolled in a plan
and is relatively happy with it can rationally decide that it is not
worth the cost of analysis (and thus just stick with her current plan),
while someone who is not yet enrolled is more likely to decide that the
analysis is worthwhile.  This explanation can be readily modeled in our
framework.  An agent's type can be taken to be a description of the
alternatives.  A TM decides how many alternatives to analyze.   There is a
cost to analyzing an alternative, and we require that the decision made
be among the alternatives analyzed or the status quo.  (We assume that
the status quo has already been analyzed, through experience.)  If the
status quo already offers an acceptable return, then a rational agent
may well decide not to analyze any new alternatives.  Interestingly, Samuelson
and Zeckhauser found that, in some cases, the status quo bias is even
more pronounced when there are more alternatives.  We can capture this
phenomenon if we assume that, for example, that there is an initial cost
to analyzing, and the initial cost itself depends in part on how many
alternatives there are to analyze (so that it is more expensive to
analyze only three alternatives if there are five alternatives
altogether than if there only three alternatives).  This would be
reasonable if there is some setup cost in order to start the analysis,
and the setup depends on the number of items to be analyzed. 
}
\exam

\section{Value of computational information}\label{sec:voci}

\subsection{Value of information: a review}
Before talking about value of computational information, we briefly
review value of information.  Consider a standard decision problem.
To deal with value of information, we consider a partition of the state
space $S$.   The question is what it
would be worth to the DM to find out which cell in the partition the true
state is in.  (Think of the cells in the partition as corresponding to the
possible realizations of a random variable $X$, and the value of information
as corresponding to the value of learning the actual realization of
$X$.)  Of course, the value may depend on the DM's type $t$.
To compute the value of information, we compute the
expected expected utility of the best action given type $t$ conditional
on receiving the 
information, and compare it to the expected utility of the best action
for type $t$
before finding out the information.  We talk
about ``expected expected utility'' here because we need to take into
account how likely the DM is to discover that he is in a particular cell.

\xam 
{\rm 
Suppose that an investor can buy either a stock or bond.  There are
two states of the world, $s_1$ and $s_2$, and a single type $t_0$.  A
priori, the investor thinks 
$s_1$ has probability $2/3$ and $s_2$ has probability $1/3$.  Buying the
bond gives him a guaranteed utility 
of 1 (in both $s_1$ and $s_2$).  In state $s_1$, buying the stock gives
a utility of 3; in state $s_2$, buying the stock gives a utility of $-4$.
Clearly, a priori, buying the stock has an expected utility of $2/3$, so
buying the bond has a higher expected utility.  What is the value of
learning the true state (which corresponds to the partition $\{\{s_1\},
\{s_2\}\}$)?  Clearly if the true state is $s_1$, buying the stock is the
best action, and has (expected) utility 3; in state $s_2$, buying the
bond is the best action, and has expected utility 1.  Thus, the expected
expected utility of 
the information is $(2/3) 3 + (1/3) 1 = 7/3$ (since with
probability $2/3$ the DM expects to learn that it is state $s_1$ and
with probability $1/3$ the DM expects to learn that it is 
$s_2$),  and so the value of information is $7/3 - 1 = 4/3$.
}
\exam

We leave it to the reader to write the obvious formal definition of
value of information in type $t$.

\subsection{Value of computational information}

In our framework, it is easy to model the value of computational
information: it is just a special case of value of information.
Formally, given a standard decision problem $(S,T,A,\Pr,u)$, we must
first extend it to a computational decision 
problem $(S',T,A,\Pr,\M,\C,\out,u')$.  $\M$ is some 
appropriate set of TMs; each TM in $\M$ outputs an action in $A$ given
an element of $S' \times T$.   As discussed 
in Section~\ref{sec:framework}, we need a richer state space to capture
the DM's uncertainty regarding the output of the TM and the running time
of the TM chosen.  We can take $S'$ to have the form $S \times
S''$, where $s'' \in S''$ determines the running time and output of each
TM $M \in \M$. Similarly, $u'((s,s''),t_0,M((s,s''),t_0), \C((s,s''),t_0,M))$
depends on $u(s,M((s,s''),t))$ and $\C((s,s''),t,M)$.  (For example, we
can assume that $u'((s,s''),t_0,M'((s,s''),t_0), \C((s,s''),t_0,M)) =
u(s,M(s,t)) - 
\C((s,s''),t,M)$, but we do not require this.) 

In this setting, value of computational information 
essentially becomes a special case of value of information. 
The only difference is that since the machine set $\M$ might be
infinite, there might not exist a machine with maximal expected
utility. So, instead of comparing the expected utilities of the best
machines (before and after receiving the information), we compare the
\emph{supremum} of the expected utilities of machine $M \in \M$ (before
and after receiving the information). 
%
More precisely, given a partition $Q$ of the state of nature, for every
cell $q\in Q$, let $\Pr_{q}$ denote the distribution $\Pr$ conditioned
on event that the state of nature is part of the cell $q$. 
and let the random variable ${\bf q}(s,t)$ denote the cell of $s$. 
The value of computational information (of learning what cell $q\in 
Q$ the state of nature is in) is 
\begin{equation}\label{eq:voci}
E_{\Pr}\left[\sup_{M \in \M} E_{\Pr_{\bf q}} \left[u'_M\right]\right] -
\sup_{M \in \M} E_{\Pr} \left[u'_M\right]. 
\end{equation}
That is, on the left-hand side, we compute the expected expected utility
by summing $\Pr(s,t) \sup_{M \in \M} E_{\Pr_{{\bf q}(s,t)}}[u'_M]$ over
all pairs $(s,t) \in S' \times T$.
Effectively, this means that the DM chooses the best TM for each cell,
after being informed what the cell is.  We discuss this issue in more
detail in Section~\ref{sec:vcon}.

Using this formalism, we can consider the value of 
learning that a particular TM $M$
is a ``good'' algorithm for the problem at hand (i.e., either learning
that it always gives the correct answer, or always runs quickly), since
this is just an event, just like learning the value of some random
variable $X$ is an event in a standard decision problem.  In a
computational decision problem,
the DM has a prior probability on $M$ being good, and
can compute the expected increase in utility resulting from
learning that $M$ is good.

\commentout{
the DM's uncertainty regarding the complexity of algorithms and the
``goodness'' of algorithms.  The primality example in the previous
section serves to make clear both types of uncertainty.   Given a TM
$M$, the DM may be uncertain about whether $\complex(s,t,M)$ is 0 or 1,
that is, whether $M$ runs in less than $2^{20}$ steps on input $t$.
Similarly, the DM may be uncertain as to whether $f(s,t,M)$ is 0, 1, or
2;  that is, the DM may be uncertain both about $M$'s output on input
$t$, and (if that output is 0 or 1), whether that is a correct
determination of $t$'s primality.  

Note that, in the primality example,  whether $M$'s output
on a given input $t$ is correct is a fact of mathematics; nevertheless,
it certainly seems reasonable that the DM should be uncertain about it.
To model this uncertainty, we assume that the DM is uncertain about the
complexity function $\complex$ and the property function $f$.  
Formally, given a decision problem $\D = (S,T,A, \Pr,\M,\complex, u)$, 
where $u$ is dependent on $f$ and $u'$, 
let $\complexs_{S, T,\M}$ and $\bF_{S, T,\M}$ be the set of all
complexity and property functions, respectively, from $S
\times T \times \M$ to $\IN$.   
An \emph{extended decision problem based on $\D$} is a tuple
$\D' = (S,T,A,\Pr^* M,u')$, 
where $\Pr^*$ is a probability on $S^* = S \times T \times
\complexs_{S,T,\M} 
\times \F_{S,T,\M}$ such that $\Pr^*$ marginalized to   
$S \times T$ is $\Pr$.  Intuitively, 
in an extended decision problem, the DM is assumed to know $u'$ and
$\Pr^*$, but not $f$ and $\complex$.  The DM's
uncertainty on $S \times 
T$ is extended to uncertainty regarding how the complexity function and
property function works.  Thus, $\Pr^*(s,t,\complex',f')$
describes the DM's probability that the state is
$s$, his type will be $t$, the complexity function is given by
$\complex'$, and the property description function is given by
$f'$.  The expected utility of a choice $M$ given $\D'$ is  
$\sum_{(s,t,f',\complex') \in S^*} \Pr^*(s,t,\complex',f')
u'(f'(s,t,M),\complex'(s,t,M))$.   $S^*$ is quite a rich space for the
DM to have beliefs about, but it seems necessary in order to consider
the value of computational information.  The following examples serve to
illustrate the richness and need for $S^*$. [[WE SHOULD PROBABLY SAY MORE
ABOUT THIS.]] 
}

\xam\label{pd3}  
{\rm
Consider the primality-testing problem from
Example~\ref{pd1}, viewed as a computational decision problem
$(S,T,A,\Pr,\M,\complex, \out,u')$.  Given the utility function, for
simplicity, we restrict $\M$ to to be a finite set of TMs that all halt 
within $2^{20}$ steps.  Thus, the DM is certain of the complexity of all
TMs in $\M$, and it is 0.  On the other hand, the DM can still be uncertain
about the output of a TM, and of the ``goodness'' of the output.
For example, if $M$ is a TM that halts after one step and outputs 0, the
DM may be certain that $M$'s output is 0, but be uncertain as to the
``goodness'' of its output.  Of course, such an algorithm might still be
worth using: if the agent places a high prior probability on the input
not being prime (which would be the case if the input was chosen
uniformly at random among all numbers less than $2^{40}$), then the
expected utility of answering 0 for all inputs is quite high.  A yet
better algorithm would be to use some naive test for primality, run it
for $2^{20}$ steps, and return 0 unless the algorithm says that the
number is prime.  The DM can then ask what the value is of learning
whether a specific TM $M$ is good (i.e., returns the correct answer for
all inputs).  This depends on the DM's prior probability that $M$ is
good; but if it is low, then the value of information is also low.
Finally, we can ask the value of being told a good algorithm (assume
that the DM is certain that there is a good algorithm, which always
returns the right answer in less than $2^{20}$ steps, but doesn't know
which it is).  This amounts to learning the value of a random variable
$X$ whose range is a subset of $\M$, where $X=M$ only if $M$ is a good TM.  
Clearly, after learning this information, the DM's expected expected
utility will be 10 (no matter what he learns, his expected utility will
be 10).  The value of this information depends on the expected utility
of the DM's best current algorithm.  Note that if the DM believes that
the input is chosen uniformly at random, then the expected utility of
even the simple algorithm that returns 0 no matter what is close to 10.
On the other hand, if the DM believes that the input is chosen so that
primes and non-primes are equally likely, the best algorithm is unlikely
to have expected utility much higher than 1 (the best strategy is likely
to involve testing whether the number is prime, outputting the answer if the
tests  reveal whether the number is prime within $2^{20}$ steps, and
outputting 2 otherwise).  In this case, the value of this information
would be close to 9.
}
\exam

\xam\label{pd4} 
{\rm
Consider the number-in-the-safe example, viewed as a
computational 
decision problem $\D = (S,T,A,\Pr,\M,\complex,\out, u')$.
Recall that the state space $S$ has the form $(s_1,s_2,s_3)$,
where $s_1$ is the number in the safe, $s_2$ is the combination of
the safe, and $s_3$ models the DM's uncertainty regarding the output of
TMs and their running time.  There is only a single type, so we can 
take $T=\{t_0\}$.  We have the obvious uniform probability
on the first two components of $S$.  Again, we restrict $\M$ to
algorithms that halt within $2^{20}$ steps.  If it takes one time unit
to test a particular combination, and the DM believes that 
the best approach is to generate some sequence of $2^{20}$ combinations
and test them, then it is clear that the DM believes that the expected
utility of this approach is $2^{-20}(1,000,000)$.  Learning the first 20
digits makes the problem feasible, and thus results in an expected
expected utility of $1,000,000$ (no matter which 20 digits are the right
ones, the expected utility is $1,000,000$),
and so has a high value of information.  
}
\exam

\commentout{
We need to be careful in defining an appropriate partition of the state
space in the primality example.  There may be several good algorithms
for primality.  What one is learned?  The choice of algorithm learned
may give more information than just the algorithm (see \cite{GH02} for
more discussion of these issues.).  For simplicity, we assume some
enumeration $M_1, M_2, \ldots$ of TMs, and that the DM learns the least
TM in the enumeration that is a good primality-testing algorithm.  Thus,
the information corresponds to a partition according to the algorithm
learned, where the cell where $M_i$ is learned consists of all these
states $(s_0,t,\complex',f')$ such that $\complex(s_0,t',M_i) = 0$ for all
$t'$, $f'(s_0,t',M_i) = 1$ for all $t'$, and for all $j < i$, there exists
a $t_j$ such that either $\complex(s_0,t_j,M_j) \ne 0$ or
$f'(s_0,t_j,M_j) \ne 1$.  

If the DM learns that $M_i$ is a good algorithm for primality testing,
then the DM believes that using $M_i$ will given an expected 
utility of 10.   
What the expected utility is of the best action without the
information depends on both the probability on $T$ and the DM's beliefs
regarding the quality of algorithms before getting the information.  For
example, if the DM believes that all inputs are equally likely, then
(given the relatively low density of primes) the strategy of
spending $2^{20}$ steps to test for primality using a standard
deterministic algorithm (recall that we have restricted to deterministic
algorithms) and then outputting 0 if the test is inconclusive will have
expected utility close to 10.  Similarly, if the DM believes that he
already has quite a good algorithm for primality, he may believe that
his expected utility is close to 10.  Note that these beliefs can all be
expressed using an appropriate probability $\Pr^*$ on the space $S^*$. 
}

\subsection{Value of conversation}\label{sec:vcon}
Recall that, for value of information, we consider how much it
is worth for a DM to find out which cell (in some partition of the state
space $S$) the true state $s$ is in. In other words, we consider the
question of how much it is worth for the DM to learn the value of $f(s)$
of some  
function $f$ on input the true state $s$. A more general setting
considers how much it is worth for a DM to interact with  
another TM $\W$ (for informant) that is running on input the true
state $s$.   
\xam 
\label{guessnumber.ex}
{\rm
Suppose a number between 1 and 100 is chosen uniformly at random.
If the DM guesses the number correctly,
he receives a utility of 100; otherwise, he receives a
utility of 0. Without any further information, the DM clearly cannot get
more than 1 in expected utility. 
But if he can sequentially ask 7 yes/no questions, he can learn the
number by 
using binary search (i.e., first asking if the number is
greater than 50; if so, asking if it is greater than 75; etc.), getting
a utility of 100.
Thus, the
value of a conversation with a machine that answers 7 yes/no questions
is 99. 
}
\exam

The \emph{value of conversation with (a TM) $\W$}
for standard decision problem
can be formalized in 
exactly the same way as value of information.
%
Formalizing computational value of conversation 
requires extending the
notion of computational decision problems to allow the DM to choose
among \emph{interactive} Turing machines $M$ (this was already done in
\cite{HP10}). 
\commentout{
Once we do this, we 
may
also allow a complexity
function $\complex$ to depend not only on the machine $M$ and
its type $t$, but also on the messages the DM receives (as well as the
random coins it tosses, if it randomizes). 
}
We omit the formal definition of an
interactive Turing machine (see, for example, \cite{goldreich01}); roughly
speaking, the machines use a special tape where the message to be sent
is placed and another tape where a message to be received is written.
We assume that the DM chooses a TM $M$.  $M$ then proceeds in two
phases.  First there is a communication phase, where $M$ converses with the
informant $\W$; then, after the communication phase is over, $M$ chooses
an action for the underlying decision problem.
Note that what an interactive TM does (that is, the message it sends or
the action it takes after the communication phase is over) can depend on its
input, the history of messages received, and the random coins it
tosses (if it randomizes).  

When considering an interactive TM $M$, we
assume that the complexity function $\complex$ depends not
only on the machine $M$ and its type $t$, but also on the messages that
the DM receives, and its random coin tosses. 
More precisely, we define the \emph{view} of an interactive machine $M$ 
to be a string $t;h;r$ in $\bit^*;\bit^*;\bit^*$,
where $t$ is the part of the type actually read by $M$, $r$ is a
finite bitstring representing the string of random bits actually used,
and $h$ is a finite sequence of messages received and read.   
If $v = t;h;r$, we take $M(v)$ to be the output of $M$ given the view.
(Note that $M(v)$ is either a message or an action in the underlying decision
problem, if the conversation phase is over.)
We now consider output functions 
$\out: {\bf M} \times S \times \bit^* \rightarrow \IN$, where ${\bf M}$
denotes a set of (interactive) Turing Machines, and let $\out(M,s,v)$
describe what the DM thinks the output of the machine $M$ is, given the
view $v$, if the state of nature is $s$. Analogously, we now consider
complexity functions $\complex: {\bf M} \times S \times \bit^*
\rightarrow \IN$, and let $\complex(M,s,v)$ describe the complexity of
the machine $M$ given the view $v$ if the state of nature is $s$.  

When running with $M$, $\W$ gets as input the actual state $s$ (we want
to allow for the possibility that $\W$ has access to some featuers of
the world that $M$ does not).  That means that the state $s$ is playing a
double role here; it is used both to capture the fact that $M$ is
interacting (in part) with nature, and may get some feedback from
nature, and to model the DM's uncertainty about the world.
To formalize the computational value of conversation with $\W$,
let the random variable ${\bf view}^{\W,M}(s,t,r_{\W},r_M)$ denote the 
view of the DM in state $s$ 
at the end of the communication phase
when communicating with $\W$
(running on input $s$ with random tape $r_{\W}$) if the DM uses
the machine $M$ (running on input $t$ with random tape $r_M$). 
We assume that ${\bf view}^{\W,M}(s,t,r_{\W},r_M)$ is generated by
computing the messages sent by $M$ and $\W$ at each step using $\out$;
that is, $M$'s first message is $\out(M,s,v_0)$, where $v_0$ is $M$'s
initial view $t;\< \,\>; r_M'$, where $\<\,\>$ denotes the empty
history, and $r_M'$ is a prefix of $r_M$, $M$'s sequence of random bits
(however much randomness $M$ used to determine its first message);
similarly, $\W$'s first message is $\out(\W,s,v_1)$, where $v_1 = s;
\<m_0\>; r_\W'$, $r_\W'$ is a finite 
prefix of $r_\W$, and $m_0$ is the
first message sent by $M$; and so on.   This means that $M$'s beliefs
about the sequence of messages sent is determined by his beliefs about
the individual messages sent in all circumstances.%
\footnote{We can allow for $M$'s beliefs about the sequence of messages
sent to be independent of his beliefs about individual messages, at the
price of complicating the framework.}

Let $\Pr^+$ denote the distribution on $S \times T \times (\{0,1\}^\infty)^2$ that is the
product of $\Pr$ and the uniform distribution on pairs of random
strings.    
%
For each pair $(\W,M)$ of interactive TMs,
we consider the random variable $u'_{\W,M}$ defined on $S\times T \times
(\{0,1\}^\infty)^2)$ by taking $u'_{\W,M}(s,t,r_{\W},r_M) =
u'(s,t,\out(M,s,v),\complex(M,s,v))$, where $v = {\bf
view}^{\W,M}(s,t,r_{\W},r_M)$. 
That is, $u'_{\W,M}(s,t,r_{\W},r_M)$ describes the utility of the
actions that result when $M$ converses with $\W$ in state $s$ given
input $t$ and random tape $r_\W$ for $\W$ and $r_M$ for $M$, taking the
complexity of the interaction into account.
The expected utility of $M$ when communicating with $\W$ is
$E_{\Pr^+}[u'_{\W,M}]$. 

The computational value of conversation with $\W$ is now defined as
\begin{equation}\label{eq:vcon}
\sup_{M \in \M}  E_{\Pr^+}\left [ u'_{\W,M}\right] - \sup_{M \in \M}
E_{\Pr^+} \left[u'_{\bot,M}\right], 
\end{equation}
where $\bot$ is the ``silent'' machine that sends no messages.  
That is, we compare the expected utility of best machine communicating
with $\W$ and the expected utility of the best machine that runs in
isolation (i.e., is communicating with $\bot$). 

There is a subtlety in this definition that is worth emphasizing. 
In general, when defining determining the best choice of TM, we must ask
whether it is reasonable to assume that the TM knows it's input.  That
is, is the choice of TM being made before the DM knows the input, or
after? 
For example, in the primality-testing problem of Example~\ref{pd1}, does
the DM choose a TM before knowing what number is or after.  The answer
to this question has no impact if we do not take complexity into
account, but it has a major impact if we do consider complexity.
Clearly, if we know what the input $n$ is, we can choose a TM that is
likely to give the right answer for $M$.  There is clearly a very
efficient TM that gives the right answer for a specific input $n$; it is
the constant-time TM that just says ``yes'' if $n$ is prime, or the
constant-time TM that just says ``no'' if $n$ is not prime.  Of course,
if there is uncertainty as to the quality of the TM, the DM may be
uncertain as to what utility he gets with each choice.  But the
complexity is guaranteed to be low.  
On the other hand, if the choice of TM must be made before the TM knows
the input, even if the DM understands the quality of the TM chosen,
there may be no efficient TM that does well for all possible inputs.  

Whether 
it is appropriate
to assume that the TM is chosen before or after the DM knows the
input depends on the application.  For the most part, in \cite{HP10}, we
implicitly
assumed that the choice was made before the DM knew the input; this
seemed reasonable for the applications of that paper.  Here, in the
definition of value of computational information, we implicitly assumed
that the DM 
chose the best TM \emph{after} learning the cell $q$ (but before
learning the input $t$).  We could also 
have computed the value of computational information under the
assumption that the TM had to be chosen before discovering $q$.
This would have amounted to putting the sup outside the scope of the
$E_{\Pr}$ in Equation (\ref{eq:voci}); this would have given
\begin{equation}\label{eq:voci1}
\sup_{M \in \M} E_{\Pr}\left[E_{\Pr_{\bf q}} \left[u'_M\right]\right] -
\sup_{M \in \M} E_{\Pr} \left[u'_M\right]. 
\end{equation}
Here we are implicitly assuming that the TM $M$ chosen takes the cell
${\bf q}(s,t)$ as an input; moreover, the TM ``understands'' that the
``right'' thing to do with ${\bf q}(s,t)$ is to condition (and thus, to
compute the expectation using $\Pr_{\bf q}$).  Again, it is possible to
allow more generality---the TM does not have to condition;  the
definition of computational value of of conversation implicitly allows this.
While (\ref{eq:voci1}) is a perfectly sensible definition, it seems less
appropriate 
when considering value of information, where a DM might be willing and
able to devote a great deal of computation to a problem after getting
information (although there may well be cases where (\ref{eq:voci1}) is
indeed more appropriate than (\ref{eq:voci})).  

By way of contrast, in (\ref{eq:vcon}), we are implicitly assuming that
the DM must choose 
the interactive TM \emph{before} learning the conversation; he does not
get to choose a different one for each conversation.  
We are evaluating the value of conversation with $\W$, rather than the
value of a particular conversation with $\W$.
This is why we do not consider the expected expected utility of the
best algorithm 
after receiving the information, but rather consider the expected
utility of ``communicating, interpreting, and finally acting''. 
Intuitively, we are assuming that a DM must choose a TM to interpret and
make use of the information gleaned from the conversation; we want to
take the cost of doing this interpretation into account, by choosing a
TM that is able to interpret all possible computations.

We could in principle define a notion of value of 
{\em particular conversations} with $\W$, rather than the value of conversing
with $\W$, by 
assuming that the DM chooses one TM that decides how to converse with
$\W$, and then, after the conversation, chooses the best TM to take
advantage of that particular conversation.  Thus, at the second step,
the TM chosen would depend on the conversation.  Formally, this amounts
to having 
another sup inside the scope of $E_{\Pr^+}$, but this seems less
appropriate here. 

If we do not take the cost of computation into account, whether we learn
the conversation before or after making the choice of TM is
irrelevant. Indeed, the 
value of conversation can be
viewed as a special case of value of information: for each
``conversation-strategy'' $\sigma$ for the DM, simply consider the value
of receiving a transcript of the conversation between $\W(s)$ and
$\sigma(t)$ (where $t$ is the type of the DM). The value of conversation
with $\W$ is then simply the maximum value of information over all
conversation strategies $\sigma$. 
By way of contrast, we cannot reduce computational value of conversation
to value of information.  If there is a computational cost associated with
computing the messages to send to $\W$, the value of a conversation is no
longer just the maximum value of information. 

\xam 
{\rm
Consider the guess-the-number decision problem from
Example \ref{guessnumber.ex} again.  What is the value of a conversation
with 
an informant $\W$ that 
picks two large primes $p$ and $q$, and sends the product
$N=pq$ to the DM? If the DM manages to factor $N$, $\W$ sends the DM the
number chosen;
otherwise $\W$ simply aborts. Clearly, the value of information in the
``best'' conversation is 99 (the DM learns the number and gets a
utility of 100). However, to implement this conversation requires the DM
to factor large number. If computation is costly and 
factoring is hard (as is widely believed), it might not be worth it for
the DM to attempt to factor the numbers. Thus, the value of
conversation with $\W$ would be 0 (or close to 0). 
}
\exam

\subsection{Value of conversation and zero knowledge}
The notion of a \emph{zero-knowledge proof} \cite{GMR} is one of the
central notions 
in cryptography. Intuitively, a zero-knowledge proof allows an
agent (called the \emph{prover}) to convince another agent (called the
\emph{verifier}) of the validity of some statement $x$, without revealing
any 
additional information. For instance, using a zero-knowledge proof, 
a prover can convince a verifier that a number $N$ is the product of 2
primes, without actually revealing the primes. 
The zero-knowledge requirement is formalized using the so-called
\emph{simulation paradigm}.  Roughly speaking, a proof $(P,V)$
(consisting of a strategy $P$ for the prover, and a strategy $V$ for the
verifier) is said to be 
\emph{perfect zero knowledge} if, 
for every 
verifier strategy $\tilde{V}$, there exists a 
simulator $S$ that can 
reconstruct the verifier's view of the interaction with the
prover
with only a polynomial overhead in runtime.%
\footnote{Technically, what is reconstructed is a distribution over
views, since 
both the prover and the verifier may randomize.}  
Note that the simulator is running in isolation and, in particular, is
not allowed to interact with the prover.  
Thus, intuitively, in a zero-knowledge proof, the verifier receives
only messages from the prover that it could have efficiently 
generated on its own by running the simulator $S$. 
The notion of \emph{precise zero-knowledge} \cite{MP06} 
aims at more precisely quantifying the knowledge gained by the
verifier. Intuitively, a zero-knowledge proof 
of a statement $x$ 
has precision $p$ if any
view that the verifier receives in time $t$ after talking to the prover
can be reconstructed by the simulator (i.e., without the help of the
prover) in time $p(|x|,t)$. 
(There is nothing special about time
here; we can also consider precision with respect more general
complexity measures.)

As we now show, there is a tight connection between the value of
conversation for computational decision problems and zero knowledge. 
To explain the ideas, we first need to introduce a new notion, which
should be of independent interest: 
\emph{value of computational speedup}.  

Computers get faster and faster. How much is it
worth for a DM to get a faster computer? To formalize this, we say that
a complexity function $\complex'$ is \emph{at most a $p$-speedup} of the
complexity function $\complex$ if, for   
all machines $M$, types $t$, and states $s$, $
\complex'(M,s,t) \le \complex(M,s,t) \le  p(\complex'(M,s,t))$. 
Intuitively, if $p$ is a constant, the value of a $p$-computational
speedup for a DM measures
how much it is worth for the DM to change to a 
machine that runs $p$ times faster than his current machine.
More precisely, 
the \emph{value of a $p$-speedup} in a computational decision
problem $\D=(S,T,A,\Pr,\M,\complex,\out, u')$ is the difference between the
maximum expected utility of the DM in $\D$ 
and the maximum expected utility in any decision problem $\D'$ that is
identical to $D$ except that the complexity function in $\D'$ is
$\complex'$, where $\complex'$ is at most a $p$-speedup of $\complex$. 

We now 
\shortv{sketch}
\fullv{present}
the connection between zero-knowledge and value of
conversation. 
Given a language $L$, an {\em objective} complexity function $\complex:
{\bf M} \times T  \rightarrow \IN$ (one that does not depend on the
state of nature), and length  parameter $n$, 
let $\D^{\complex}_{L,n}$ denote the class of computational decision
problems $\D=(S,T,A,\Pr,\complex', \out, \M, u)$, where $\M$ is the set of
interactive Turing machines, $S \subseteq \{0,1\}^n$,
types in $T$ have the form $x;t'$, where $x \in S$ and $t' \in
\{0,1\}^*$, and $\Pr$ is  
such that $\Pr(s,t) > 0$ only if $s = x$, $t = x;t'$, and $x \in L$
(so that the DM knows $x$ and that $x \in L$). 
We also require that (1) the DM does not have any uncertainty about the
output and the complexity functions: for all $M,s,t$, $\out(M,s,t)=M(t)$
(so the DM knows the correct outputs of all machines) and
$\complex'(M,s,t)=\complex(M,t)$ (so the DM knows the complexities of
all machines); and (2) 
$\D$ is \emph{monotone in complexity}:
for all types $t \in T$, actions $a \in A$, and complexities $c \le c'$,  
$u(t,a,c) \geq u(t,a,c')$; that is, the DM never prefers to compute more. 
\shortv{
In the full paper we prove the following.}
\fullv{
We prove the following theorem in Appendix \ref{zkproof.sec}.}
\begin{theorem}\label{zk.thm}
If $(P,V)$ is a zero-knowledge proof system for the
language $L$ with precision $p(\cdot,\cdot)$ with respect to the
complexity function $\complex$, then for all $n \in N$ and all
computational decision problem $\D \in \D^{\complex}_{L,n}$, the value
of conversation with $P$ in $\D$ is no higher than the value of a
$p(n,\cdot)$-computational speedup in $\D$. 
\end{theorem}
Thus, intuitively, 
if the DM is not uncertain about the complexities and the outputs of
machines, the value of participating in a zero-knowledge proof is 
never higher than the value of (appropriately) upgrading computers.

\section{Discussion and Related Work}\label{sec:discuss}
We have introduced a formal framework for decision making that
explicitly takes into account the cost of computation.  Doing so
requires taking 
into account the uncertainty that a DM may have about the running time
of an algorithm, and the quality of its output.  The framework allows us
to provide formal decision-theoretic solutions to well-known
observations such as the status-quo bias and belief polarization.

Of course, we are far from the first to recognize that decision making
requires computation---computation for knowledge acquisition and for
inference.  Nor are we the first to suggest that the costs for such
computation should be explicitly reflected in the utility function.  
Horvitz \citeyear{Hor87} credits Good \citeyear{Good52} for being the
first to explicitly integrate the costs of computation into a framework
of normative rationality.  For example, Good points out that ``less good
methods may therefore sometimes be preferred'' (for computational
reasons).  In a sequence of papers (see, for example, \cite{Hor87,Hor01}
and the references therein), Horvitz continues this theme, investigating
various policies that trade off deliberation and action, taking into
account computation costs.  The framework presented here could be used
to provide formal underpinnings to Horvitz's work. 

In terms of next steps, we have considered only one-shot decision
problems here.  It would be very interesting to extend this framework
to sequential decision problems.  Moreover, we have assumed that agents can
compute the probability of (or, at least, are willing to assign a
probability to) events like ``TM $M$ will halt in 10,000 steps'' or
``the output of TM $M$ solves the problem I am interested in on this
input''.  Of course, calculating such probabilities itself involves
computation.  Similarly, calculating utilities may involve computation;
although the utility was easy to compute in the simple examples we gave,
this is certainly not the case in general.  It would be  
relatively straightforward to extend our framework so that the TMs
computed probabilities and utilities, as well as actions.   
\shortv{
In this
setting, it may make sense to allow for a more general
representation of uncertainty.  That is, an agent may start with a set
of probabilities rather than a single probability, and may then refine
that set (perhaps to a single probability) over time.  Similarly, an agent may
start with a set of possible utilities, rather than a single utility.

Once we allow sets of probabilities and utilities, we need to reconsider
how to define the notion of ``optimal choice''.  We could, for example,
use the maxmin expected utility approach of Gilboa 
and Schmeidler \citeyear{GS1989}, associating with each action the
worst-case expected utility (over all probability distributions and
utility functions considered possible) and choose the action with the
best worst-case expected utility; other approaches may also be
reasonable.  }
However, once we do this, we need to think about what counts as an
``optimal'' decision if the DM does not have a probability and utility,
or has a probability only on a coarse space.
An alternative approach might be to allow the set of TMs that the DM
considers possible to increase (at some computational cost), but assume
that DM has all the relevant probabilistic information about the TMs
that it can choose among.  As this discussion should make clear, there
is much fascinating research to be done in this area.

\shortv{
Considering sequential decision-making also allows us to 
examine \emph{consistency} of decisions.  Taking cost of computation
into account may make decisions appear consistent that are not
consistent without taking cost of computation into account.  
As this discussion should make clear, there is much
fascinating research to be done in this area.
}

\bibliographystyle{chicagor}
\bibliography{z,joe}

\fullv{
\appendix


\section{Precise Zero Knowledge}
\label{zkproof.sec}
We recall the definition of \emph{precise zero knowledge} \cite{MP06,P06},
which is an extension of the notion of zero knowledge \cite{GMR}.
%
We start by informally recalling the notion of an interactive proof; see
\cite{GMR,goldreich01} for more details. 
An \emph{interactive proof} 
is a pair $(P,V)$ of protocols, where $P$ is used by the \emph{prover} 
and $V$ is used by the \emph{verifier}.  The pair
$(P,V)$ is required to satisfy two properties for every input $x$:
\emph{completeness}---if $x \in L$, then after running $(P,V)$ with
input $x$, the verifier is convinced 
that $x \in L$
with
probability 1, and \emph{soundness}---if $x \notin L$, then 
the verifier running $V$
should 
reject with overwhelming probability, no matter what strategy 
$P'$
the prover uses. 

Roughly speaking, zero-knowledge proofs are interactive proofs where the
verifier does 
not learn anything beyond the fact that $x\in L$ from the interaction
with the prover. 
This is formalized by requiring that for every strategy $V'$ of the verifier,
there exists a \emph{simulator} strategy $S$ that, given a
string $x$ and whatever auxiliary information $z$ the verifier may have,
can 
reconstruct the view of the verifier when running $V'$ on input $(x,z)$
with the prover running $P$ on input $x$.  
(Note that in the definition of zero knowledge, the verifier may have 
some information $z$ as well as the input string $x$.
The information $z$ is often called the \emph{auxiliary input}, and
models any 
prior knowledge the verifier has about $x$.)
 
The traditional notion of zero knowledge
requires only that 
the \emph{worst-case} complexity (size and running-time) of
the simulator
$S$ is 
polynomially related to that of $V'$.
Precise zero knowledge additionally
requires that the complexity of the simulator $S$ ``respects''
the complexity of the verifier's protocol $V'$ in an ``execution-by-execution''
fashion: a zero-knowledge proof is said to have precision $p(n,t)$ if the
running time of the simulator $S$ (on input an instance $x \in \{0,1\}^n$)  
be bounded by $p(n,t)$ whenever $S$ outputs a view in which the
running time of $V'$ is $t$. 
We generalize the definition of \cite{MP06}, which considers only
the running time as a complexity measure by allowing arbitrary
(objective) complexity measures $\complex : {\bf M} \times S \times T
\rightarrow \IN$. 
In our generalization, instead of  bounding only the complexity of $S$,
we require that the complexity of the combined machine $V'(S(\cdot)$ is
bounded by $p(n,t)$ whenever $S$ outputs a view in which the
running time of $V'$ is $t$. (Note that if we take the complexity of
$V'$ to be its running time, then this change is inconsequential:
running $V'$ on 
the output of $S$ adds only $t$ computational steps, since $t$ is the
running time of $V'$ on the view output by $S$.)  

\begin{definition}[Perfect Precise $\ZK$] \label{pzk.def} Let $L$ be a
language, $(P,V)$ an interactive proof system for $L$, $\complex$ a
complexity function, and $p:N\times N \times N \rightarrow N$ a
monotonically increasing function. We say that $(P,V)$ is 
\emph{perfect $\ZK$} with $\complex$-\emph{precision} $p$ if, for
every 
interactive Turing machine $V'$, there exists a probabilistic algorithm
$S$  such that the 
following two conditions hold: \begin{enumerate}
\item [1.] For all $x\in L$ and $z\in \{ 0,1\}^*$, the probability that
the verifier 
receives the view $v$ in an interaction between $P$ on input $x$ and
$V'$ on input $(x,z)$ is identical to the probability that $S(x,z)$
outputs $v$. 
\item [2.] For all $x \in L$ and $z\in \{0,1\}^*$, and all
sufficiently long $r\in \{0,1\}^*$,  
$\complex(V'(S),(x,z,r)) \leq p (|x|, \complex(V',S(x,z,r))$.
(Here $r$ is a random string, which is needed to determine the output of
$S$, since $S$ may randomize.)
\end{enumerate} \end{definition}

\subsection{The proof}
\begin{theorem} [Theorem \ref{zk.thm} restated]
If $(P,V)$ is a zero-knowledge proof system for the
language $L$ with precision $p(\cdot,\cdot)$ with respect to the
complexity function $\complex$ then, for all $n \in N$ and all
computational decision problems $\D \in \D^{\complex}_{L,n}$, the value
of conversation with $P$ in $\D$ is no higher than the value of a
$p(n,\cdot)$-computational speedup in $\D$. 
\end{theorem}
\prf
Recall that, given a language $L$, complexity function $\complex$, and length
parameter $n$, 
$\D^{\complex}_{L,n}$ denotes the class of computational decision
problems $\D=(S,T,A,\Pr,\M,\complex', \out, u')$, where $\M$ is the set
of 
interactive Turing machines, $S \subseteq \{0,1\}^n$, 
types in $T$ have the form $x;t'$, where $x \in S$ and $t' \in
\{0,1\}^*$, and $\Pr$ is 
such that $\Pr(s,t) > 0$ only if $s = x$, $t = x;t'$, and $x \in L$,
for all $M,s,t$, $\out(M,s,t)=M(t)$, 
$\complex'(M,s,t)=\complex(M,t)$,
and $u$ is monotone in complexity.
Consider some language $L$, $n \in N$, decision problem $\D \in
\D^{\complex}_{L,n}$, and an interactive TM $M$. Assume $M$ gets
expected utility $d$ in $\D$ after interacting with $P$. 
We show that there exists a decision problem $\D'$ and a machine $M'$
such that $\D'$ is a $p(n,\cdot)$-computational speedup of $\D$ and $M'$
has expected utility $d' \geq d$ in $\D'$. This implies that the value
of conversation with $P$ in $\D$ is no higher than the value of a
$p(n,\cdot)$-computational speedup in $\D$. 

Let $S$ be the zero-knowledge simulator for $M$, and let $M'(v) = M(S(v))$ (i.e., $M'$ first runs $S$ and then runs $M$ on the output of $S$). Let $\tilde{\complex}$ be defined identically to $\complex$ except that for every $v$, 
$\tilde{\complex}(M(S),v) = \complex(M,S(v))$.
Consider the decision problem $\tilde{D}$ that is identical to $\D$ except that the complexity function is $\tilde{\complex}$.
Since, for every input $(x,z)$, the output of $S(x,z)$ is identically
distributed to the view of $M(x,z)$ in an interaction with $P(x)$, and
since  
$\tilde{\complex}(M(S),v) = \complex(M,S(v))$, it follows that
$U_{\tilde{D}}(M') = d$.
Let ${\complex'}$ be defined identically to $\complex$ except that for
every $v$, ${\complex}(M(S)),v) = \complex(M,S(v))$ if
$\complex(M,S(v))\leq \complex(M(S(\cdot)),v)$ and
$\complex(M(S),v)$ otherwise. 
Since utility is monotone in complexity, we have that 
$$U_{\D'}(M') \geq U_{\tilde{\D}} = d,$$
which concludes the proof.
\eprf
} 

\end{document}